\pdfoutput=1

\documentclass[article,superscriptaddress,aps,preprintnumbers,amsmath,amssymb,prd,nofootinbib,twocolumn]{revtex4}

\usepackage{epsfig}
\usepackage{amssymb}
\usepackage{subfigure}
\usepackage{cancel}
\usepackage{textcomp}
\usepackage{calc}
\usepackage{graphicx}
\usepackage{dcolumn}
\usepackage{color}
\usepackage{xcolor}
\usepackage{pifont}
\usepackage{slashed}
\usepackage{fdsymbol}
\usepackage{orcidlink}
\usepackage{braket}
\usepackage[normalem]{ulem}
\usepackage{diagbox}
\usepackage{float}

\usepackage{booktabs,amsmath,caption,dcolumn}
\newcolumntype{d}[1]{D..{#1}}

\hypersetup{
colorlinks   = true, 
urlcolor     = red, 
linkcolor    = blue, 
citecolor   = blue 
}
\usepackage[capitalise]{cleveref}
\usepackage[normalem]{ulem} 
\newcommand{\nn}{\nonumber}

\begin{document}


\title{Memory of a Random Walk: Astrometric deflections from gravitational wave memory accumulation over cosmological scales}

\author{T\"{o}re Boybeyi\orcidlink{0000-0002-2877-1507}} 
\email{boybe001@umn.edu}
\affiliation{School of Physics and Astronomy, University of Minnesota, 55455 MN, USA
}

\author{Vuk Mandic\orcidlink{0000-0001-6333-8621}} 
\email{vuk@umn.edu}
\affiliation{School of Physics and Astronomy, University of Minnesota, 55455 MN, USA
}

\author{Alexandros Papageorgiou\orcidlink{0000-0002-2736-3026}} 
\email{papageorgiou.hep@gmail.com}
\affiliation{Particle Theory and Cosmology Group, Center for Theoretical Physics of the Universe, Institute for Basic Science (IBS), 34126 Daejeon, Republic of Korea
}
\affiliation{Instituto de F\'isica T\'eorica, UAM-CSIC, 28049, Madrid, Spain}


\begin{abstract}
We study the impact of gravitational wave memory on the distribution of far away light sources in the sky. For the first time we compute the built up of small, but permanent tensor distortions of the metric over cosmological time-scales using realistic models of compact binary coalescences (CBCs) whose rate of occurrence is extrapolated at $z\sim {\cal O}(1)$. This allows for a consistent computation of the random-walk like evolution of gravitational wave memory which, in turn, is used to estimate the overall shape and magnitude of astrometric deflections of far away sources of light. We find that for pulsar or quasar proper motions, the near-Earth contribution to  the astrometric deflections dominates the result and the deflection is analogous to a stochastic gravitational wave memory background that is generally subdominant to the primary stochastic gravitational wave background. We find that this contribution can be within the reach of future surveys such as \textit{Theia}. Finally, we also study the deviation of the presently observed angular distribution of quasars from perfect isotropy, which arises from the slow build-up of gravitational wave memory over the entire history of the universe. In this case, we find that astrometric deflections depend on the entire light trajectory from the source to the Earth, yielding a quadruple pattern whose magnitude is unlikely to be within reach of the next generation of astrometric surveys due to shot noise and cosmic variance limitations. 
\end{abstract}


\date{\today}


\maketitle

\section{Introduction}
\label{sec:introduction}

In the past few years we have witnessed a flurry of exciting discoveries in gravitational wave (GW) astrophysics. Advanced LIGO (aLIGO)~\cite{aLIGO} and Advanced Virgo (aVirgo)~\cite{aVirgo} observations of mergers of binary black holes (BBH)~\cite{GW150914}, binary neutron stars (BNS)~\cite{GW170817}, and black hole-neutron star (BHNS) systems~\cite{LIGO_NSBH}, as well as pulsar timing array measurements of the stochastic gravitational wave background~\cite{NANOGrav:2023gor,EPTA:2023fyk,Reardon:2023gzh,Xu:2023wog} have established GW astrophysics on firm grounds. They have also paved the way for a host of new tests of fundamental physics such as tests of general relativity~\cite{GW150914TGR,O3TGR}, constraints on the nuclear matter equation of state in neutron stars~\cite{GW170817}, new measurements of the Hubble constant $H_0$~\cite{GW170817H0, O3H0}, estimates of the stochastic gravitational wave background (SGWB) from BBH \cite{GW150914stoch, O3ratepop} and BNS~\cite{GW170817stoch} systems, constraints on dark matter from primordial BBHs (PBBH)~\cite{PBBH}, early universe scenarios \cite{EPTA:2023xxk} and more. 

Despite achieving nearly 100 detections of GWs from BBH systems to date~\cite{GWTC3} at least one well established prediction of general relativity remains elusive: gravitational wave memory (GWM). At its core, GWM is a phenomenon which entails a permanent change in the displacements of freely falling masses due to the passage of a GW. Observing GWM with terrestrial GW detectors such as aLIGO and aVirgo is challenging due to the relatively low-frequency nature of GWM, typically below the sensitive band of terrestrial detectors. However, it is possible that GWM could be detected using an ensemble of ($\sim 2000$) BBH mergers with terrestrial detectors~\cite{GWM_LIGO,Grant:2022bla}. On the other hand, space based detectors such as LISA have a higher potential for detecting GWM~\cite{GWM_Favata,Ghosh:2023rbe} and PTA experiments have already placed strong upper limits on the GWM strain~\cite{Agazie:2023eig}. The linear form of GWM has been known since the 1970's~\cite{Braginsky:1985vlg,Zeldovich:1974gvh,Braginsky1987} with the nonlinear form discovered in the 1990's~\cite{Blanchet:1992br,PhysRevD.45.520}. Since then, our understanding and modeling of GWM has improved further~\cite{Bieri:2013hqa,Bieri:2013ada,Tolish:2014bka,Tolish:2014oda,Tolish:2016ggo,Garfinkle:2017fre,Bieri:2017vni,Bieri:2020zki,Jokela:2022rhk} (see also~\cite{GWM_Favata} for a review) and deeper relationships between GWM, soft theorems and BMS (Bondi-Metzner-Sachs) transformations (infrared triangle) have been unveiled~\cite{Strominger:2014pwa}.

Most of the literature on GWM has focused on investigating the properties of GWM arising from individual events such as from BBH mergers~\cite{GWM_Colm} or cosmic strings~\cite{GWM_Mairi} and these studies roughly suggest that the GWM amplitude is expected to be an order of magnitude lower than the magnitude of the parent GW signal. Only recently several authors have begun investigating the phenomenology of stacking multiple GWM signals~\cite{Allen:2019hnd,GWM_Zhao,Cao:2022can}. Unlike the previous studies, \textit{the purpose of the present work is to investigate the properties of the accumulated GWM in a patch of space arising from all possible sources over cosmological time scales and their impact on astrometric deflections}. The basis of our work is the fact that a single GW burst with memory will cause a transient (primary GW) and permanent (GWM) change in the metric within a patch of space under consideration. Both effects consist of ``stretching" one direction while ``shrinking" another. However, the transient, although a stronger effect, will eventually decay away while the weaker contribution will leave a permanent metric distortion in the patch of space under consideration. One can then consider the build-up of GWM over cosmological time scales by adding the GWM contributions from all GW sources in the history of the universe, modeling each contribution as a step-like, permanent, quadrupolar distortion of space time. 
The effect described above can be conceptualized as a Brownian motion of ``stretching" and ``shrinking" of spacetime, with the mean metric distortion averaging to zero but with a standard deviation that scales as the square root of the number of GW sources/events. The sheer number of events over the entire cosmological history of the universe make it so this GWM accumulation could lead to deviations from isotropic expansion~\cite{Cao:2022can}, cause redshift and deflections of light from distance sources such as quasars~\cite{Madison:2020xhh}, and even cause distortions of the CMB~\cite{Madison:2020atu}. Our work aims to establish the foundations for a more systematic understanding of the phenomenology of cumulative GWM and to that end we specifically focus on astrometric deflections as a first step.

When embarking on the calculation of accumulated GWM, one should in principle sum over all possible GW sources throughout the history of the universe. This would include contributions of SGWB from inflation~\cite{grishchuk,barkana,starob,turner,peloso_parviol,seto,Guzzetti:2016mkm}, early universe phase transitions~\cite{witten,hogan,turnerwilczek,kosowsky,kamionkowski,apreda,caprini1,binetruy,caprini2}, additional ``stiff" phases of cosmological evolution~\cite{boylebuonanno}, cosmic strings~\cite{caldwellallen,DV1,DV2,cosmstrpaper,olmez1,siemens,ringeval,olum,O1cosmstr,O3cosmstr,Weir:2017wfa} and GWs produced by recent astrophysical processes~\cite{phinney,regfrei,zhu_cbc,marassi_cbc,rosado,regman,wu_cbc,GW150914stoch,GW170817stoch,cutler,bonazzola,marassi_magnetar,owen,rotatingNS,barmodes2,barmodes3,regman,wu_mag,regimbau_review,SNe,marassi_cc, firststars,buonanno_cc,crocker1,crocker2, finkel,LIGOScientific:2021nrg}. Performing a complete calculation that will include all possible sources is a daunting task as it would involve modeling sources of GWs with very large uncertainties such as the cosmological SGWB (from inflation, cosmic strings and phase transitions). We limit the scope of our analysis to contributions arising from mergers of compact objects such as BBHs, supermassive BBHs (SBBH), and primordial BBHs (PBBH) as the modeling of GWM strain for such sources in terms of the parameters of the primary GWs is well understood. In the case of BBHs of stellar origin, the local merger rate is well measured by aLIGO and aVirgo observations, and the redshift evolution of the BBH population can be modeled well using star formation rate~\cite{KAGRA:2021duu,LIGOScientific:2021djp}. In the case of SBBH and PBBH, both the local merger rate and its extrapolation to high redshifts are highly uncertain but constrained within certain limits \cite{Enoki:2004ew,Raidal:2017mfl}. Keeping all this in mind, our calculation of accumulated GWM should be understood as a lower limit since we are neglecting a plethora of other GW sources that would increase the overall GWM.

Given a model for the rate of accumulation of memory in the patch of space of interest, perhaps the most immediate and compelling phenomenological application is its potential impact on astrometric surveys of far away sources such as those performed already by \textit{Gaia} \cite{gaia2016gaia,brown2021gaia,lindegren2021gaia,vallenari2023gaia,babusiaux2023gaia} or the planned Telescope for Habitable Exoplanets and Interstellar/Intergalactic Astronomy (\textit{Theia}) \cite{boehm2017theia,Malbet:2022lll,garcia2021exploring}. As is well established~\cite{Madison:2020xhh}, a single gravitational wave burst with memory causes a deflection in the angular distribution of light sources in the sky. This deflection consists of two terms, coined the `Earth' and `Star' terms. For the primary GW, the Earth term dominates while the Star term is strongly suppressed. On the other hand, for GWM the Earth and Star terms can be comparable and the permanent deflections induced by GWM may depend on the entire history of the universe. Our primary aim is to provide a realistic calculation for the memory build-up over cosmological scales and hence for the total astrometric deflection build-up which in principle could be imprinted in the statistics of the distant light sources proper motions or angular distributions. The physics of GWM in the context of astrometric deflections is complementary to the astrometric deflections by primary SGWB that have been studied extensively in the literature \cite{Moore:2017ity,Klioner:2017asb,Mihaylov:2018uqm,Mihaylov:2019lft,jaraba2023stochastic}.

This work is organized as follows: In Section \ref{sec:accumulation} we outline the formalism that allows us to compute the accumulation of GWM over cosmological scales for different types of black hole binaries. In Section \ref{sec:merger_rates} we provide the details needed for accurately modeling the binary merger rates and mass distributions for the BBH, SBBH, and PBBH populations. In Section \ref{sec:results} we show the results of our numerical analysis, computing the GWM for different populations of binaries. 
In Section \ref{sec:astro-def} we discuss detection prospects for cumulative GWM focusing on astrometric deflections, and we conclude in Section \ref{sec:conclusions}. We will set $c=1$.

\section{GWM accumulation}
\label{sec:accumulation}

As outlined above, one key difference between the primary GW and GWM is that the former is transient while the latter is permanent. As a result, we expect the latter to accumulate over time.
We focus our attention to a small patch of space whose memory accumulation, over cosmological time scales, we want to compute. The patch of space is sufficiently small compared to the distance to the sources of GWs, so that we can assume the GWM to be homogeneous within it and the wavefronts to be planar. In this context, every GW signal arriving from a random direction and with random polarization will induce a small, but permanent, change in the metric in the chosen patch. Of course, such changes can have differing amplitudes and signs, implying that cumulative GWM evolves in a random walk fashion with zero mean and a standard deviation that scales as the square root of the number of GW sources. This cumulative random walk process results in small permanent metric fluctuations (anisotropy) from patch to patch~\cite{Cao:2022can}, which may impact trajectories of light from distant sources~\cite{Madison:2020xhh,Madison:2020atu} and cause aparent deflections of these sources on the sky. 

The basic ingredient necessary for our calculation is the local merger rate per comoving volume in the source frame which we denote as $R_X(z)$ where $X={\rm BBH/SBBH/PBBH}$ is an index that specifies the source of GWs. Explicit form of $R_X(z)$ depends on the type of BBH population we consider, and we will specify it below for the three cases we consider. We then convert the merger rate to the observer frame by dividing by redshift and we further multiply by the differential comoving volume $dV/dz$ to obtain the observer merger rate
\begin{align}\label{eq:mergerratetoday}
    R_{X,z}(z)  =  \frac{R_X(z)}{(1+z)}\frac{dV}{dz}= \frac{R_X(z)}{(1+z)}\frac{4\pi D^2(z)}{H_0 E(z)} 
\end{align}
where $D(z)$ is the comoving distance to the source and $E(z)=\sqrt{\Omega_m(1+z)^3+\Omega_\Lambda}$ accounts for the expansion of the universe with the standard energy densities of dark matter, $\Omega_m = 0.3$, and dark energy, $\Omega_\Lambda = 0.7$. Alternatively, $H(z)=H_0 E(z)$ is the Hubble constant as a function of time/redshift. We then generalize this rate into the merger rate as observed at a redshift $z_o$:
\begin{equation}\label{mrate}
    \tilde{R}_{X,z}\left(z_o,z_{so}\right)
    =\frac{R_X(z_s)}{(1+z_{so})}\frac{4\pi D^2(z_o,z_s)}{H_0 E(z_{so})}
\end{equation}
where we have defined $z_{so}$ as the redshift of the source at $z_s$ relative to the observing redshift $z_o$:
\begin{equation}
    1+z_{so}=\frac{1+z_s}{1+z_o}.
\end{equation}

\begin{figure}
\includegraphics[width=0.48\textwidth,angle=0]{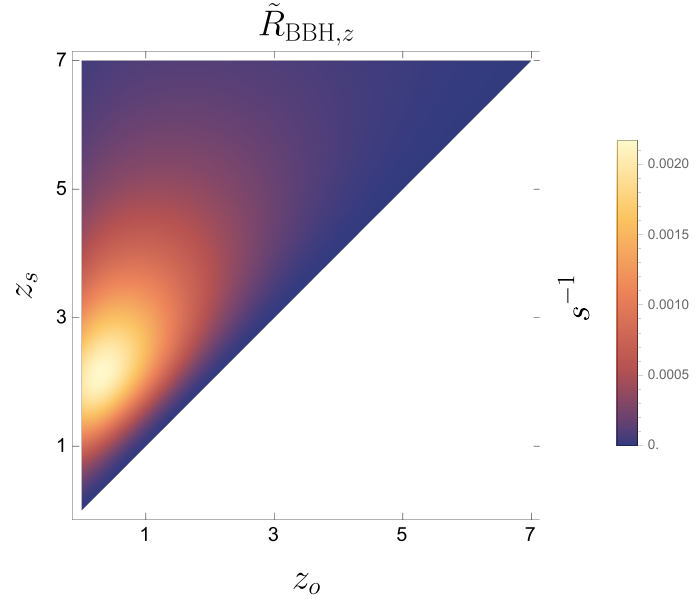}
    \caption{Cumulative GWM calculation integrates all GWM contributions from all sources over the entire history of the universe, as depicted by the shaded triangular region. Here we display the merger-rate for stellar BBH population defined in Eq. \eqref{mrate}.}
    \label{fig:cartoon}
\end{figure}

In order to compute the total GWM in the given patch accumulated over the entire history, one needs to integrate the GWM contributions generated by GWs arriving at each observer redshift $z_o$, which in turn depend on all GW sources at redshifts $z_{s}>z_o$. The integration region is depicted in Figure \ref{fig:cartoon}.

In addition to the merger rate, one also needs the GWM strain for each individual event. 
We will employ the so called \textit{Step Model}, $h(t) = \Theta(t-t_m)  h_{+,{\rm mem}}(z_{so},M,\theta)$, where $\Theta(t)$ denotes the Heaviside step function, $t_m$ is the time of the merger, and $h_{+,{\rm mem}}$ denotes the amplitude of the memory signal. For compact binary mergers, the plus polarization is typically dominant and is well approximated by \cite{Jokela:2022rhk}
\begin{eqnarray}
    h_{+,{\rm mem}}(z_{so},M,\theta)=\frac{G M}{144 D(z_{o},z_s)}\zeta(\theta)
    \label{Eq:memstrain}
\end{eqnarray}
where $D(z_{o},z_s)$ is the comoving distance to the GW source to an observer at $z=z_o$, $\theta$ is the angle between the binary's orbital plane and the binary's location in the sky, and $M$ is the mass of a component black hole of the binary system, assuming the two component black holes have the same mass. The angular factor is given in \cite{Jokela:2022rhk} as $\zeta(\theta)=\sqrt{2}\left(17+\cos^2\theta\right)\sin^2 \theta$.

Note that for simplicity we multiplied the result due to the inspiraling phase by a factor of 2 to model the contribution of the ringdown phase. This is a rough approximation that is sufficient to get the correct order of magnitude and it significantly simplifies the expressions we use below. Note also that we assumed the masses of the two compact objects to be the same for simplicity. 

Finally, in addition to the above, we need a mass distribution from which we draw a random value for the mass of the black holes for each event. We denote the probability distribution as $g_X(M)$ and assume it to be independent of the redshift for simplicity. 

The elements outlined above can be used to sum the total GWM over cosmological scales. Specifically, we assume that strain $h_+$ occurs at a rate $\tilde{R}_{X,z}\left(z_o,z_{so}\right)$, with a uniformly distributed inclination angle $\theta$ and direction in the sky and with a mass drawn from distribution $g_X(M)$. 

We use the formalism in \cite{abbott2016gw150914} to express SGWB in terms of merger rate as observed at a redshift $z_o$,
\begin{align}
    \Omega_{\rm mem} (f,z_o) = \frac{f}{\rho_c H_0} \int^{\infty}_0 dz_{so} \frac{R_{X}(z_s) \braket{\frac{dE_{\rm mem}}{df}(f_s,z_{so})}}{(1+z_{so}) \sqrt{\Omega_{\rm \Lambda}+\Omega_{\rm M}(1+z_{so})^3}} 
\end{align}
where $\rho_c=3H^2_0/8\pi G$ and $f_s = f(1+z_s)$ is frequency of the GW evaluated at the source. The ensemble average denoted by brackets means averaging over the mass and angular distributions given by $\braket{\dots} = \int d\Omega \sin(\theta) \int dM g_X(M)$.
In the \textit{Step Model} and using Eq. \ref{Eq:memstrain}, we have
\begin{align}
    h(f) = \frac{1}{2\pi i f} \frac{ G M}{144 D(z_o,z_s)} \zeta(\theta) 
\end{align}
up to a delta function at $f=0$ which is not relevant for the following derivations and \cite{phinney2001practical}
\begin{align}
    \frac{dE_{\rm mem}}{df}(f_s,z_{so}) = \frac{\pi^2}{G}  D^2(z_o,z_s) (1+z_s)^2 f^2 |h(f(1+z_s))|^2
\end{align}
Combining these,
\begin{align}\label{Eq:GWMOmega}
   \Omega_{\rm mem} (f,z_o) &= f \ \frac{2\pi G^2}{3H^3_0} \frac{\sigma^2_{ \theta}}{(144^2)}  \int dM M^2 g_X(M) \nn \\&\times \int^{\infty}_0 dz_{so} \frac{R_X(z_s) }{(1+z_{so})\sqrt{\Omega_{\rm \Lambda}+\Omega_{\rm M}(1+z_{so})^3}}
\end{align}
where
\begin{align}\label{angfactor}
    \sigma^2_{\theta} = \int_0^{\pi}d\theta\, \sin(\theta) \zeta(\theta)^2 \simeq 627
\end{align}
denoting the angular contribution. The lower limit of the $d{z_{so}}$ integral is chosen to be zero, which implies that the patch under consideration, whose memory can be understood as homogeneous, is infinitesimal in size. If one wants to consider finite size patches, the lower limit should be adjusted to ensure that the sources of gravitational waves are much farther away than the characteristic patch size.

In the next section we compute the $\Omega_{\rm mem} (f,z_o)$ for different populations of binaries and then use it to numerically compute the total memory accumulation over cosmological scales.  

\section{Sources of GWM and merger rates}
\label{sec:merger_rates}

We dedicate this section to elucidating the modeling of the merger rates of black hole binaries and their extrapolation to high redshifts for the three populations under consideration:  BBH, PBBH, and SBBH. In particular, for each of the three populations we specify the model and parameters for the merger rate $R_X(z)$ and the mass distribution $g_X(M)$, enabling the computation of Eq. \ref{Eq:GWMOmega}. Table \ref{table:pop} lists the ranges of relevant parameters used in the three population models. 

\begin{table}
\centering
\caption{Population Model Parameters. Here $\mathcal{N}(\mu,\sigma)$ denotes normal distribution whereas $U_{[a,b]}$ denotes the uniform distribution.}
\label{table:pop}
\begin{tabular*}{0.5\textwidth}{@{\extracolsep{\fill}} l c @{}}
\toprule[0.4pt]\toprule[0.4pt]
Parameters & Range/Value \\
\midrule[0.4pt]
\multicolumn{2}{c}{\rule{3cm}{0.4pt} BBH parameters \rule{3cm}{0.4pt}} \\
$\nu \ (\text{Gpc}^{-3} \ \text{yr}^{-1})$ & $\mathcal{N}(150,10^2)$ \\
$p$ & $\mathcal{N}(2.37,0.1^2)$ \\
$z_m$ & $\mathcal{U}_{[2.0,2.4]}$ \\
$q$ & $\mathcal{N}(1.8,0.1^2)$ \\
$m$ & $\mathcal{U}_{[2.0,2.4]}$ \\
$\Bar{M}$ & 30 $M_{\odot}$ \\
$\sigma_m$ & 5 $M_{\odot}$ \\
$M_l$ & 5 $M_{\odot}$ \\
$M_h$  & $\mathcal{U}_{[80,100]M_{\odot}}$  \\
\midrule
\multicolumn{2}{c}{\rule{3cm}{0.4pt} PBBH parameters \rule{3cm}{0.4pt}} \\
$M_c$ &  $\mathcal{U}_{[25,30]M_{\odot}}$  \\
$\sigma_M$ & $\mathcal{U}_{[0.1,0.3]M_{\odot}}$  \\
$f_{\text{sup}}$ & $10^{-3}$ \\
$\alpha$  & $\mathcal{U}_{[1.0,1.6]}$ \\
$f_{\text{PBH}}$  & $\mathcal{U}_{[0,1]}$ \\
\midrule
\multicolumn{2}{c}{\rule{3cm}{0.4pt} SMBBH parameters \rule{3cm}{0.4pt}} \\
$\log_{10} \frac{\Dot{n}_0}{(\text{Mpc}^{-3}\text{Gyr}^{-1})}$ & $\mathcal{N}(-3,1)$ \\
$\alpha_M$ & $\mathcal{U}_{[-2,2]}$ \\
$\log_{10}\frac{M_{*}}{M_{\odot}}$ & $\mathcal{U}_{[6.5,8.5]}$ \\
$\beta_z$ & $\mathcal{U}_{[0,7]}$ \\
$z_c$ & $\mathcal{U}_{[0,5]}$ \\
\bottomrule[0.4pt]
\end{tabular*}
\end{table}

\subsection{Binary Black Hole Mergers}
\label{sec:merger_rates_sub1}

\begin{figure*}
    \centering
    \includegraphics[width=0.48\linewidth,angle=0]{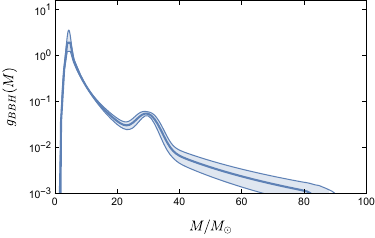}
    \includegraphics[width=0.48\linewidth,angle=0]{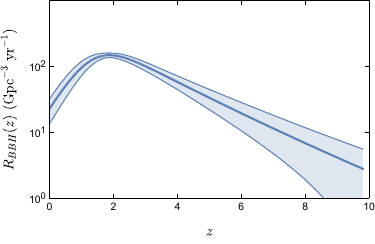}
    \caption{Mass distribution (left) and merger rate (right) for BBH using the models in Eqs. \eqref{bbhmass} and \eqref{bbhrate} with values in Table \ref{table:pop}. Bands denote the 1-sigma (68\%) confidence region, computed as described in the text.}
    \label{fig:pop1}
\end{figure*}

Binary black hole systems of stellar origin are well constrained by ground based GW detectors and their properties are relatively well understood. We assume that the BBH merger rate follows the star formation rate \cite{hernquist2003analytical}:
\begin{align}\label{bbhrate}
   R_{\rm BBH}(z)= \nu  \frac{ p e^{q(z-z_m)} }{(p-q) + q e^{p(z-z_m)}}
\end{align}
with the parameter $\nu$ normalizing the merger rate to the observed merger rate at $z=0.2$.  The peak of the $\text{SFR}$ is defined by $z_m$, while $p-q$ and $q$ define slopes of $\text{SFR}$ at high and low redshifts. We have chosen the range of these parameters (see Table \ref{table:pop}) so that they are consistent with the LIGO O3 results \cite{abbott2023population}. We draw free parameters of the BBH merger rate model from their distributions given in Table \ref{table:pop}, calculate $R_{\rm BBH}(z)$ for each parameters choice, and plot the 68\% confidence region of computed $R_{\rm BBH}(z)$ curves in Figure \ref{fig:pop1} (right).

The BBH mass distribution is assumed to follow the power-law peak model with the lower cut-off at $5 M_{\odot}$ and high cut-off around $[80,100] M_{\odot}$:
\begin{align}\label{bbhmass}
    g_{\rm BBH}(M) =& \Big((1-\lambda)\frac{M^{-m}}{\lambda_1} + \lambda \frac{e^{-(M-\Bar{M})^2/(2\sigma^2_M)}}{\lambda_2}  \Big)\nonumber\\ &\times \Theta(M_h-M)\Theta(M-M_l).
\end{align}
Here, $\lambda_1,\lambda_2$ are normalization constants of the two distribution components and $\lambda=0.08$ is the mixing factor. We assume the mass distribution does not evolve with redshift. Typical ranges of these parameters are given in Table \ref{table:pop}. Again the ranges of parameters are chosen to be in agreement with the LIGO O3 results \cite{abbott2023population}.

Similar to the merger rate, we draw free parameters of the BBH mass model from their distributions given in Table \ref{table:pop}, and calculate $g_{\rm BBH}(M)$ for each parameters choice, then plot with the corresponding 68\% confidence region on the left side of Figure \ref{fig:pop1}.

\subsection{Primordial Binary Black Hole Mergers}
The primordial black hole mass distribution is commonly assumed to be log-normal~\cite{Raidal:2018bbj,Flitter:2020bky}:
\begin{equation}
    g_{\rm {PBBH}}(M)=\frac{1}{\sqrt{2\pi}\sigma_M M}{\rm exp}\left(-\frac{\log^2(M/M_c)}{2\sigma_M^2}\right)
    \label{eq:PBHdistribution}
\end{equation}
where $M_c$ is the peak mass of $M g(M)$ and $\sigma_M$ characterizes the width of the mass function. Log-normal distribution is usually motivated by baryonic dark matter models as proposed in \cite{dolgov1993baryon}.

While the mass distribution is different from the stellar BBH case, the more important difference is in the binary merger rate as a function of redshift. The PBBH systems do not follow the evolution of stellar material, and are instead related to the evolution of dark matter halos. 
The PBBH merger rate has been studied in several works \cite{Ali-Haimoud:2017rtz,Raidal:2018bbj,Vaskonen:2019jpv,Atal:2020igj,Mukherjee:2021ags,mandic2016stochastic}. As an example, the formalism of \cite{Mukherjee:2021ags} assumes a simple power-law dependence of the merger rate with redshift:
\begin{equation} \label{eq:PBHmerger}
    R_{\rm PBBH}(z)=R_{\rm PBBH} (0) (1+z)^{\alpha},
\end{equation}
with $\alpha \sim 1.3$ for a Poisson spatial distribution of the PBBH systems~\cite{Raidal:2017mfl}. In this scenario, the local merger rate for equal mass black holes is given by \cite{Raidal:2018bbj,Clesse:2020ghq}:
\begin{equation}\label{eq:PBHtoday}
    \frac{R_{\rm PBBH}(z=0)}{{\rm Gpc}^{-3}\,{\rm yr}^{-1}}=4\times 10^5 f_{\rm sup}\,f_{\rm PBH}^{53/37}\,\left(\frac{M}{M_\odot}\right)^{-32/37}
\end{equation}
where $f_{\rm PBH}$ is the fraction of dark matter in the form of primordial black holes ($\leq 1$ by definition) and $f_{\rm sup}$ is a suppression factor that depends on the effects from other PBBHs and the matter distribution surrounding it. In \cite{Clesse:2020ghq}, the value of $f_{\rm sup} \sim 10^{-3}$ (chosen to be a benchmark number) was argued to be consistent with current LIGO observations giving consistent merger rates and also motivated by $N$-body simulations. 

\begin{figure*}
    \centering
    \includegraphics[width=0.48\linewidth,angle=0]{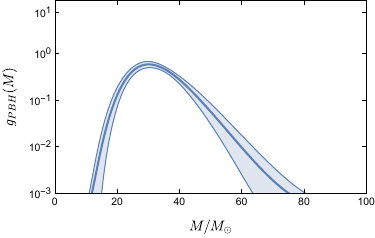}
    \includegraphics[width=0.48\linewidth,angle=0]{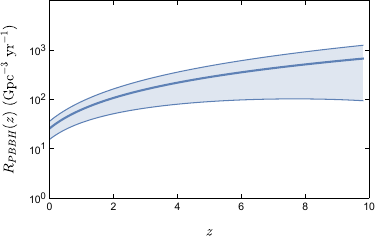}
    \caption{Mass distribution (left) and merger rate (with $M = M_{\odot}$ in \eqref{eq:PBHtoday}) (right) for PBBH using the models in Eqs. \eqref{eq:PBHdistribution}, \eqref{eq:PBHmerger} and \eqref{eq:PBHtoday} with values in Table \ref{table:pop}. Bands denote the 1-sigma (68\%) confidence region, computed as described in the text.}
    \label{fig:pop2}
\end{figure*}

Ranges for all of the free parameters of the PBBH model are chosen to be consistent with the SGWB in \cite{Mukherjee:2021ags} and are given in Table \ref{table:pop}. 68\% confidence regions for the merger rate and mass models for PBBH are calculated in the same way as for the stellar BBH model and are plotted in Figure \ref{fig:pop2}.

\subsection{Supermassive Binary Black Holes Mergers}
There are various models for SBBH mergers. The two most common models used in the literature are the so called astrophysically-informed, \cite{chen2019constraining}, ($\dim(\theta_i)=16$) and the agnostic/minimal, \cite{middleton2015astrophysical}, ($\dim(\theta_i)=5$) model. Parameters of these models are constrained by Pulsar Timing Array (PTA) experiments, for example in \cite{steinle2023implications}.

We will use the agnostic model outlined in \cite{middleton2015astrophysical,middleton2018no,steinle2023implications} as it captures the essential properties of SBBH population while retaining simplicity. In this model, merger rate is given by a Schechter function
\begin{align}\label{smbbhmerger}
   R_{\rm SBBH}(z) = \Dot{n}_0 (1+z)^{\beta_z} e^{-z/z_c}
\end{align}
with mass probability density
\begin{align}\label{smbbhmass}
     g_{\rm SBBH}(M) = \frac{1}{M} \Big(\frac{M}{10^7 M_{\odot}} \Big)^{-\alpha_M} e^{-M/M_{\star}}
\end{align}
\begin{figure*}
    \centering
    \includegraphics[width=0.48\linewidth,angle=0]{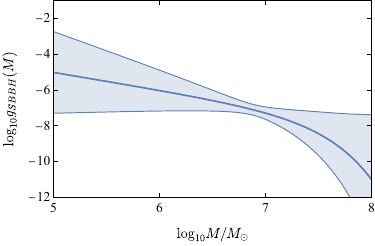}
    \includegraphics[width=0.48\linewidth,angle=0]{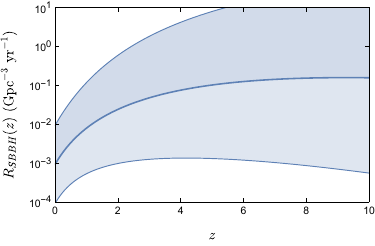}
    \caption{Mass distribution (left) and merger rate (right) for SBBH using the models in Eqs. \eqref{smbbhmerger} and \eqref{smbbhmass} with values in Table \ref{table:pop}. Bands denote the 1-sigma (68 \%) confidence region, computed as described in the text.}
    \label{fig:pop3}
\end{figure*}

Agnostic model is based on the assumption that binary systems undergo mergers in circular orbits, influenced only by radiation reaction. The density of these mergers, denoted by $\Dot{n}_0$, is measured per unit of rest-frame time and co-moving volume. For the distribution's characteristics in terms of $M$, the parameters $\alpha_M$ and $M_\star$ specify the slope and the cut-off point, respectively. Meanwhile, the distribution in $z$ is similarly defined by the parameters $\beta_z$ and $z_c$, which represent the same features for the merger rate.

The ranges of all free parameters of the SBBH model are chosen to obey the PTA constraints \cite{steinle2023implications,middleton2015astrophysical,middleton2018no} and are given in Table \ref{table:pop}. The 68\% confidence regions for the SBBH merger rate and mass models are calculated in the same way as in the previous cases and are plotted in Figure \ref{fig:pop3}.
\section{Cumulative GWM Estimates}   
\label{sec:results}

Memory accumulation can be characterized by a 1D random walk. The step size at each time is a Gaussian random variable with zero mean and a time dependent variance, $\sigma(t)$. Probability of accumulating memory strain $h$ after time $t$ is given by 
\begin{equation}
P(h) \propto \exp\left(-\frac{h^2}{2 \sigma^2(t)}\right)
\label{Eq:rwdist}
\end{equation}
We define the so called diffusion constant which is an invariant characterization for random walks by $D (t) = \Dot{\sigma} \sigma$, where the dot represents time derivative.

We note that the diffusion constant is itself a function of time in our case. The time evolution and the amplitude of the diffusion constant depend on the parameters of the population/merger rate model outlined in the previous section. Each realization of the population is a set of random parameters drawn from the distributions given in Table \ref{table:pop}. Therefore each realization yields a different curve $D(z_o)$.

We relate the diffusion constant to the stochastic gravitational wave background due to memory. The power spectral density for a random walk is given by
\begin{equation}
    S_{\text{mem}}(f) = \frac{D}{\pi^2 f^2}.
\end{equation} 
valid for frequencies lower than the merger time scale of the binaries. Combining this with the standard relationship between the strain power spectrum and the energy density of the stochastic background,
\begin{equation}
    \Omega_{\text{mem}}(f) = \frac{2 \pi^2 f^3 S_{\text{mem}}(f)}{3H^2_0}
    \label{Eq:strainomega}
\end{equation}
yields
\begin{equation}
    \Omega_{\text{mem}}(f) = \frac{2D f}{3H^2_0},
\end{equation}
This in our case generalizes to 
\begin{align}
    \Omega_{\text{mem}}(f,z_o) = \frac{2D(z_o) f}{3H^2_0}
\end{align}
where $\Omega_{\rm mem} (f,z_o)$ is given by Eq. \ref{Eq:GWMOmega}.

\begin{figure*}
  \centering  \includegraphics[width=0.48\linewidth,angle=0]{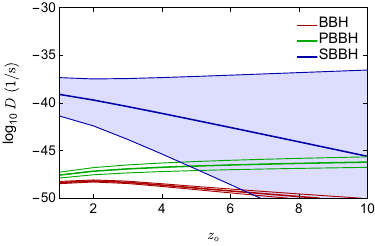}
\includegraphics[width=0.48\linewidth,angle=0]{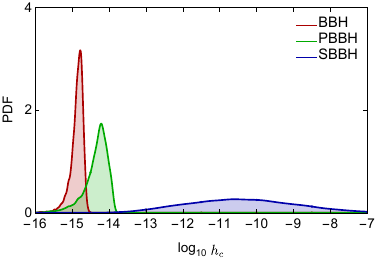}
\caption{Diffusion constant $D$ as a function of redshift along with it's 1-sigma (68 \%) confidence region (left) and corresponding diffusion length probability distribution (right).}
\label{fig:D}
\end{figure*}

On the left-side of Figure \ref{fig:D}, we show the evolution of the diffusion constant for the three GW source populations: BBH, PBBH, and SBBH. The bands of each source correspond to 1-sigma interval, i.e. the regions where the central $68\%$ of the $D(z_o)$ curves lie in. The central solid curves are the mean of the realizations. The present time diffusion constant values $D(0)$ are comparable to previous results \cite{zhao2022stochastic}, which did not investigate the redshift dependence of the memory accumulation process.
Our results, therefore, extrapolate $D(0)$ to high redshifts. 

Having the full $D(z_o)$, one can make an estimate of the typical strain accumulation $h_c$, which is given by the symmetric Gaussian distribution defined in Eq. \ref{Eq:rwdist}.  More precisely, the typical magnitude of $h_c$ is given by the diffusion length, defined as
\begin{equation}
h_c \approx 2\sqrt{\int^{z=0}_{z_o=10} D(t) dt}.    
\end{equation}
Figure \ref{fig:D} (right) shows the probability distribution of $h_c$ for different sources. We note that SBBH sources would dominate the cumulative memory effects (reaching the effective strain $10^{-11}-10^{-9}$) as compared to the BBH and PBBH models, with $\sim 4$ orders of magnitude uncertainty in the diffusion length today due to the large uncertainty in the SBBH population model.
\section{Astrometric deflections}
\label{sec:astro-def}

\begin{figure}[H]
\begin{minipage}{.5\textwidth}        \includegraphics[width=0.48\linewidth,angle=0]{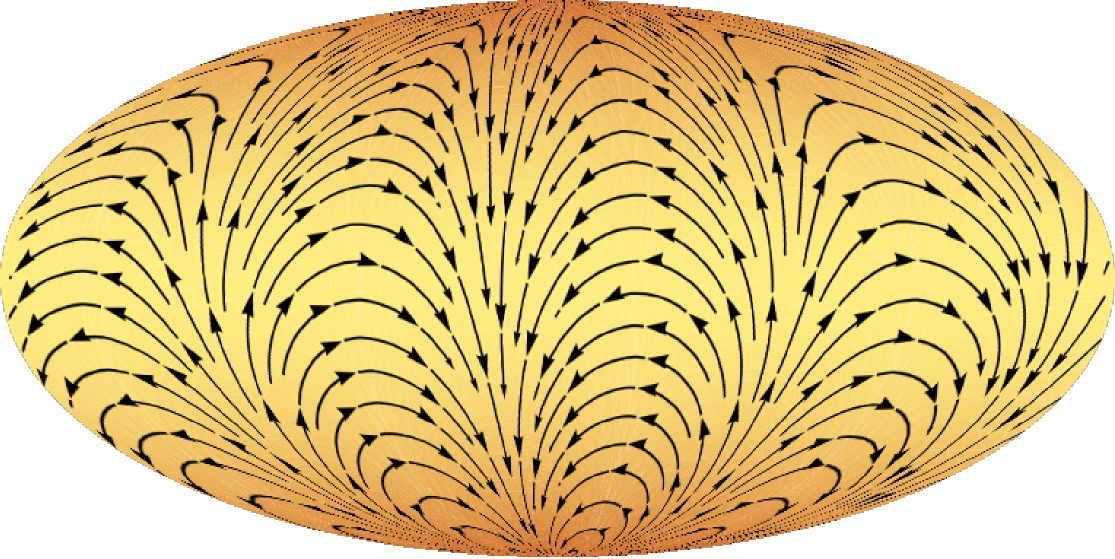}
\includegraphics[width=0.48\linewidth,angle=0]{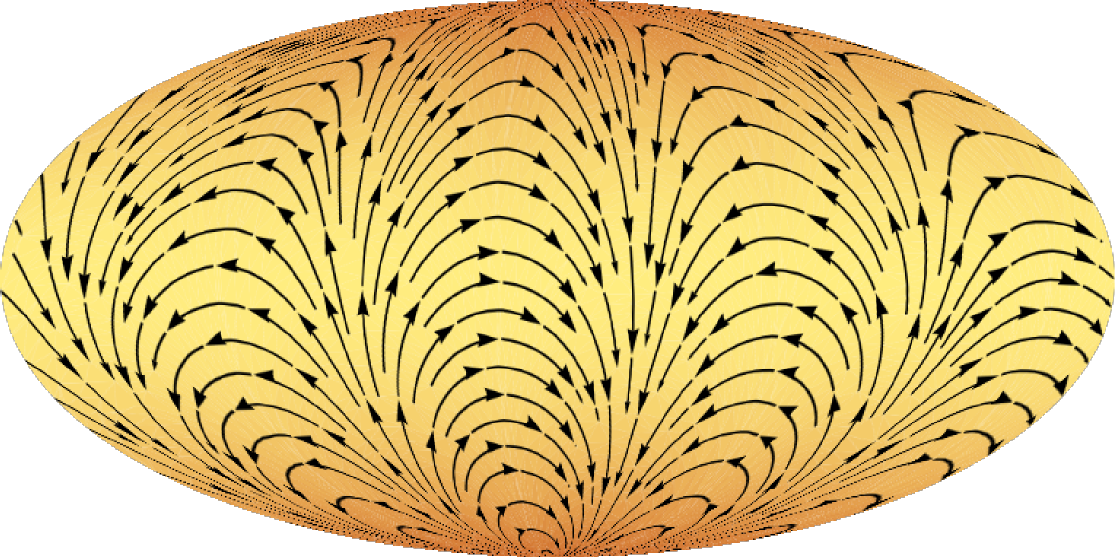}
\end{minipage}
\begin{minipage}{.5\textwidth}
\centering
\includegraphics[width=0.48\linewidth,angle=0]{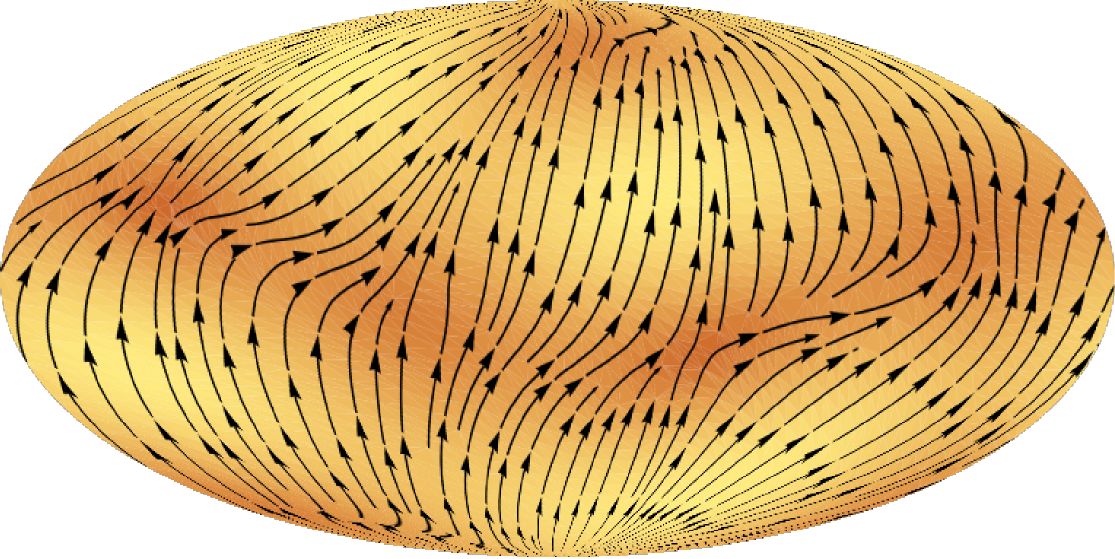}
\end{minipage}
    \caption{Top two figures depict deflection patterns of $+$ and $\times$ polarization of $\mathcal{V}^{\plus,\times}_{\bigoplus}$, respectively, for a GW propagating in the $\hat{z}$ direction. The bottom figure depicts a  realization of deflection for many such GWs uniformly distributed in direction and polarization.}
    \label{fig:def}
\end{figure}

One compelling application for the GWM accumulation derived in the preceding section is the contribution to astrometric deflections of far away objects such as pulsars or quasars. Our work is based on the treatment of astrometric deflections by GWs and GWM by Madison \cite{Madison:2020xhh} which makes use of the formalism of Book \& Flanagan \cite{book2011astrometric}. The observed angular deflection of a distant light source located at $n^i$, due to a propagating tensor perturbation can be expressed as Eq. (35) of \cite{book2011astrometric},
\begin{align}
    \delta n^i = &P^{ik} n^j \Big( -\frac{h_{jk}(0)}{2}+ \frac{p_k n_l}{2(1+\boldsymbol{p}\cdot \boldsymbol{n})}h_{jl}(0) \nonumber \\&+\frac{1}{\lambda_s}\int^{\lambda_s}_0 d\lambda \big[h_{jk}(\lambda)-\frac{p_k n_l}{2(1+\boldsymbol{p}\cdot \boldsymbol{n})}h_{jl}(\lambda)\big]  \Big)
    \label{Eq:deltan}
\end{align}
with $P^{ik}= \delta^{ik}-n^i n^k$ being the transverse-traceless projector and $h_{ij}(\lambda)=\epsilon^A_{ij}(\boldsymbol{p}) h(t-\lambda (1+\boldsymbol{p}\cdot \boldsymbol{n}))$ is the strain evaluated on the unperturbed path at an earlier time with $p^i$ denoting the wave direction. Here, $\lambda_s$ is the proper distance to the light source and the source of the tensor perturbation is assumed to be much further away than $\lambda_s$ so that the plane wavefront approximation holds. The expression coincides with the one for an expanding universe (\cite{book2011astrometric}) up to an overall scale factor correction which we take to be one since we will consider only gravitational wave sources with ${\cal{O}}(1)$ redshift. This can be simplified and subsequently split in terms of "Earth" and "Star" contributions (Eq. (5) of \cite{Madison:2020xhh}),
\begin{align}\label{def}
    \delta n^i(t) =  \mathcal{V}^{i,A}_{\bigoplus}(\boldsymbol{p}) h(t) + \mathcal{V}^{i,A}_{\bigstar}(\boldsymbol{p})\frac{H(t)-H(t-\lambda_s(1+\boldsymbol{p}\cdot\boldsymbol{n}))}{\lambda_s(1+\boldsymbol{p}\cdot \boldsymbol{n})} 
\end{align}
with
\begin{align}\label{earthdef}
   \mathcal{V}^{i,A}_{\bigoplus}(\boldsymbol{p}) =  -\frac{ n^j \epsilon^{i,A}_{j}}{2} + n^l n^j \epsilon^{A}_{jl} \Big[ \frac{ p^i + n^i}{2(1+\boldsymbol{p}\cdot \boldsymbol{n})}\Big]
\end{align}
\begin{align}
    \mathcal{V}^{i,A}_{\bigstar}(\boldsymbol{p}) =  n^j \epsilon^{i,A}_{j} -\frac{n^i n^l n^j \epsilon^{A}_{jl}}{2} - n^l n^j \epsilon^{A}_{jl} \Big[ \frac{ p^i + n^i}{2(1+\boldsymbol{p}\cdot \boldsymbol{n})}\Big]
\end{align}
where $H = \int h(\lambda) \,d\lambda$ and $A=\plus,\times$ is the polarization index. The star term encapsulates propagation effects while the Earth term captures the deflection due to the strain at the 
observer location. The deflection pattern of the Earth term $\mathcal{V}^{i,A}_{\bigoplus}$ for different polarizations are shown in Fig. \ref{fig:def}. 

One interesting limit, which will be referred to as the 'Saturation Limit', is defined as $\delta n^i_s\equiv \delta n^i(\infty)$ and corresponds to the final deflection configuration, long after the gravitational wave has passed over the local patch of space spanning the light source and observer. It is straightforward to calculate this limit from Eq. \eqref{def}---letting $h(t)=A_0 \Theta(t)$, one obtains
\begin{align}
    \delta n^i_s(\infty) = A_0 (\mathcal{V}^{i,A}_{\bigoplus}(\boldsymbol{p}) + \mathcal{V}^{i,A}_{\bigstar}(\boldsymbol{p})) = \frac{A_0}{2}P^{ij} \epsilon^{A}_{jk} n^k .
    \label{satdef}
\end{align}

Paying particular attention to the memory part of the signal, it is important to note that, we see from Eq. \eqref{def} that unlike for regular GWs whose Star term is suppressed by $\sim 1/\lambda_s$, the Star term for GWM grows as $\sim t/\lambda_s$ and gradually becomes $\mathcal{O}(h)$. This in turn implies that for GW signals which passed over the Earth in recent years, one would expect the Earth term to yield the greatest contribution to the total astrometric deflection of a far away source. 
On the other hand, for GWs that passed over the Earth long ago, the contribution of the Star term is greater---i.e. in general, it can increase even after the primary wavefront has propagated beyond the Earth. Eventually, if we consider GWs that passed over the patch of space defined by the light source and the Earth even before the light was emitted (that we observe today), the overall effect of the astrometric deflection would be precisely the Saturation Limit we defined above. In that case, the light propagated through the already perturbed metric from the moment it was emitted until it finally arrived at the Earth. 

Due to the considerations outlined above, we separate our analysis into two distinct observational scenarios. 

\begin{itemize}
    \item Observations of the proper motion of far away sources through astrometric deflections during the observation time. In such a scenario, one would compare how the location of sources of light such as pulsars or quasars change over the time scale of the observations (e.g. ${\cal O}(10)$ years). The dominant contribution to this measurement will arise from the Earth term of GWs that propagate by the Earth within the time of observation while the star term provides a small, low frequency correction. 
    \item Observations of the angular distribution of far away sources, such as quasars, based on a single time snapshot. Deviation of this distribution from perfect isotropy could then be attributed to GWM, and we will analyze the statistics of such a deviation. In this scenario, the Earth and Star terms contribute on the same order and GWs emitted throughout the history of the universe are important.
\end{itemize}

\begin{figure*}
    \centering
    \includegraphics[width=0.48\linewidth,angle=0]{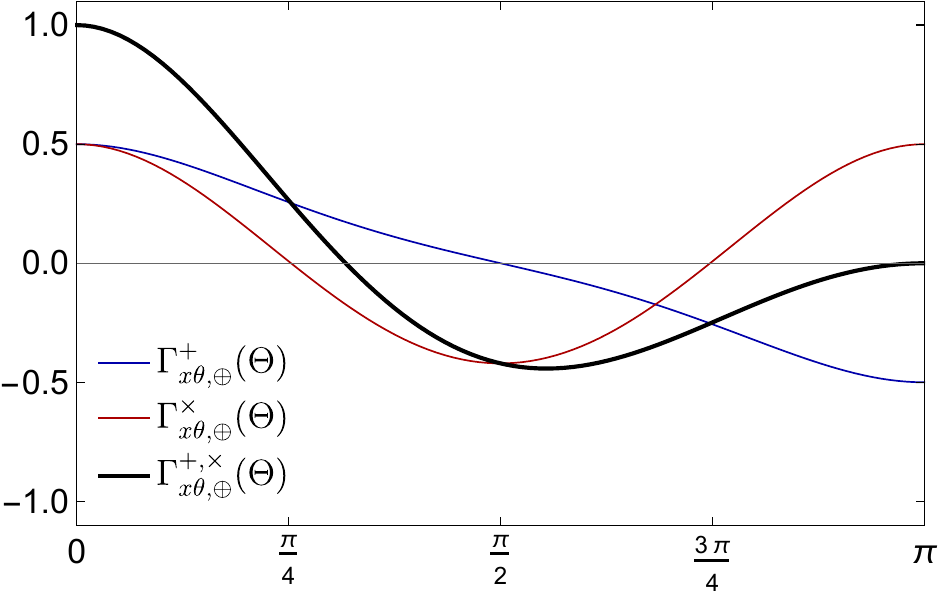}
    \includegraphics[width=0.48\linewidth,angle=0]{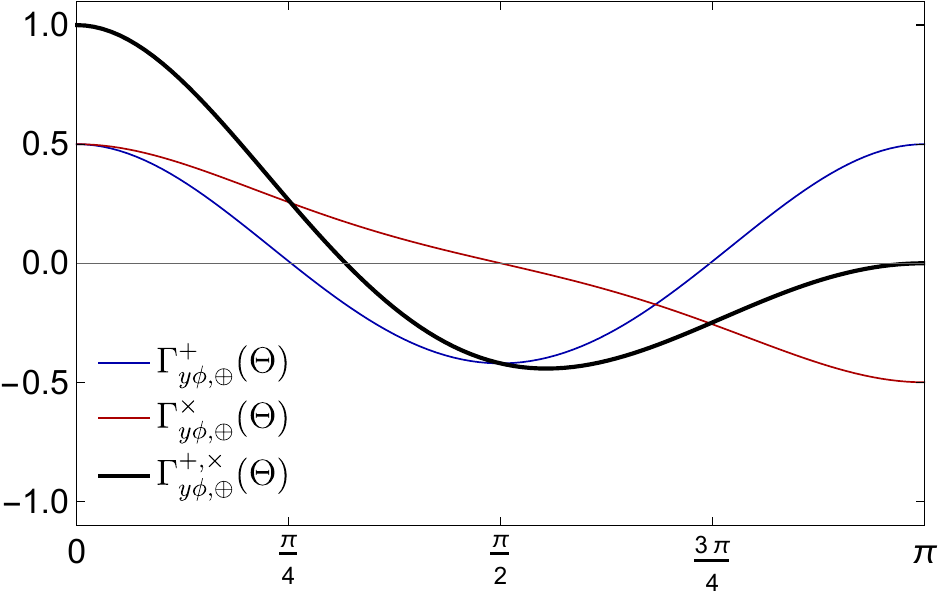}
    \includegraphics[width=0.48\linewidth,angle=0]{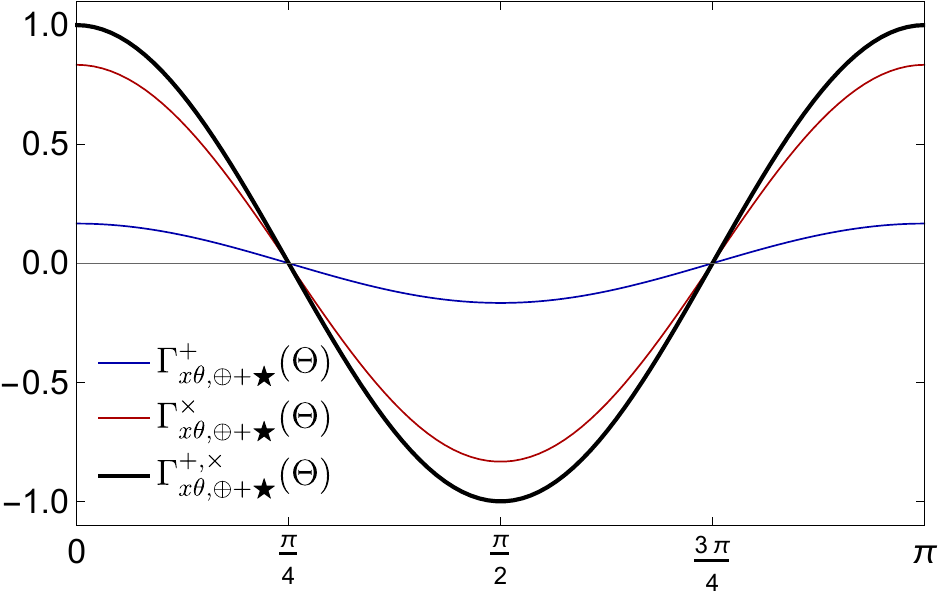}
    \includegraphics[width=0.48\linewidth,angle=0]{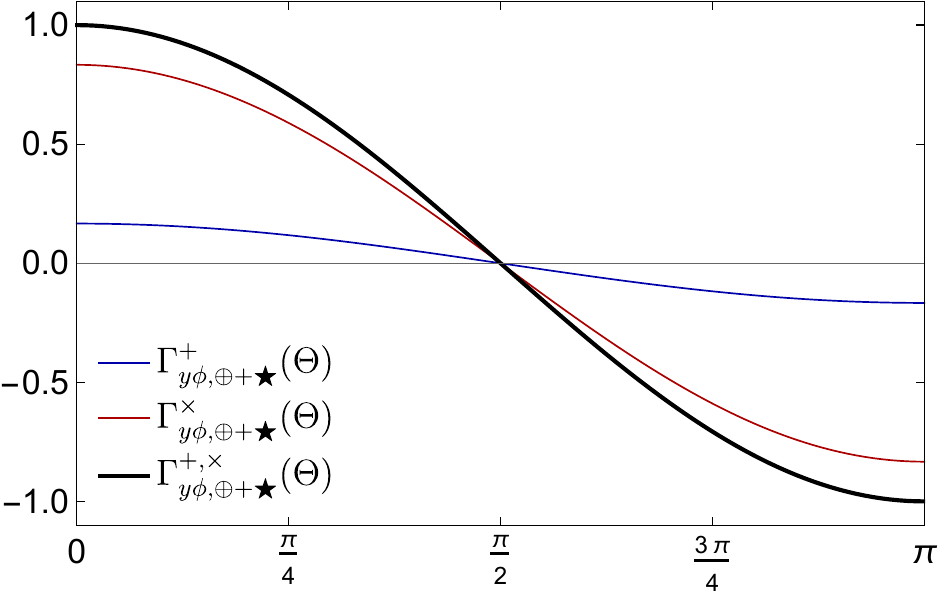}
    \caption{Correlation function for Earth term (top) given in Eq. \eqref{earthdef} and the saturation limit (bottom) given in Eq. \eqref{satdef} are shown after expansion in orthogonal basis functions as defined in Eq. \eqref{basis}. }
    \label{fig:corr}
\end{figure*}

For each of these scenarios, it is required to calculate the so-called geometric correlation function. We define it in the following way,
\begin{align}
    \Gamma^A_{ij,\bigoplus}(\boldsymbol{n},\boldsymbol{n}') = \int d^2 \boldsymbol{p}   \mathcal{V}^{i,A}_{\bigoplus}(\boldsymbol{n},\boldsymbol{p})   \big(\mathcal{V}^{j,A}_{\bigoplus}(\boldsymbol{n}',\boldsymbol{p}) \big)^{*} .
\end{align}
for the Earth term and 
\begin{align}
    \Gamma^A_{ij,\bigoplus+\bigstar}(\boldsymbol{n},\boldsymbol{n}') = \int d^2 \boldsymbol{p}   &\Big(\mathcal{V}^{i,A}_{\bigoplus}(\boldsymbol{n},\boldsymbol{p}) + \mathcal{V}^{i,A}_{\bigstar}(\boldsymbol{n},\boldsymbol{p}) \Big) \\ \times  & \Big(\mathcal{V}^{j,A}_{\bigoplus}(\boldsymbol{n}',\boldsymbol{p}) + \mathcal{V}^{i,A}_{\bigstar}(\boldsymbol{n},\boldsymbol{p})\Big)^{*} .
\end{align}
for the saturation limit. The above quantities are tensors but for practical reasons it is more useful to expand them in a suitable basis with scalar coefficients. Let us define the following basis.
\begin{align}\label{basis}
    &\boldsymbol{u}_{y} = \frac{(\boldsymbol{n}\times \boldsymbol{n}')}{\sqrt{1-(\boldsymbol{n}\cdot \boldsymbol{n}')^2}} \qquad \boldsymbol{u}_x = \boldsymbol{u}_y \times \boldsymbol{n}\\& \qquad \boldsymbol{u}_{\phi}=\boldsymbol{u}_y \qquad \boldsymbol{u}_\theta = \boldsymbol{u}_\phi \times \boldsymbol{n}'
\end{align}
The are four possible projections, for example $ \Gamma^A_{x\theta,\bigoplus} \equiv \Gamma^A_{ij} u^i_x u^j_\theta$, $ \Gamma^A_{x\theta} \equiv \Gamma^A_{ij} u^i_y u^j_\phi$ etc. The non-zero projections of $\Gamma^A_{ij,\bigoplus}$ and $\Gamma^A_{ij,\bigoplus+\bigstar}$ are shown on Fig. \ref{fig:corr} with the definition $\cos(\Theta)=\boldsymbol{n}\cdot \boldsymbol{n}^{'}$. Using these, one can reconstruct these tensor geometric two point correlations as
\begin{align}
    \label{corrdef1}&\Gamma^{A}_{{ij}_{\bigoplus}}(\Theta) = \sum_{x_i=x,y.} \sum_{\theta_j=\theta,\phi.} \Gamma^{A}_{{x_i\theta_j}_{\bigoplus}} (\Theta) u^i_{x_i} u^j_{\theta_j} \\ \label{corrdef2}
    &\Gamma^{A}_{{ij}_{\bigoplus+\bigstar}}(\Theta) = \sum_{x_i=x,y.} \sum_{\theta_j=\theta,\phi.} \Gamma^{A}_{{x_{i}\theta_{i}}_{\bigoplus+\bigstar}} (\Theta) u^{i}_{x_{i}} u^{j}_{\theta_{j}}
\end{align}

We observe that in the case of the Earth term the $\Gamma_{x\theta}$ and $\Gamma_{y\phi}$ correlations functions (with the polarizations exchanged) are identical. In the case of the Star term, this degeneracy is broken and all contributions are distinct.
As a final note, in this paper we are using conventions in \cite{Mihaylov:2018uqm} to set the basis vectors (see \cite{book2011astrometric,Mihaylov:2018uqm} for an analytical derivation of the Earth correlation function). The Star term is usually neglected in the literature as it is multiplied by $H(t)$ which saturates to a finite value for oscillatory waves and as a result the star term is suppressed by $H(t)/\lambda_s$. In the case of the memory, however, the Star term remains unsuppressed until it saturates to a final value comparable to the Earth term, as described by the saturation limit defined above.

\subsection{Impact on the proper motions of far away objects}
\label{sec:scenarioA}
We dedicate this section to analyzing the first of the two scenarios to constrain GWM accumulation. We will focus on the dominant SBBH source. As mentioned above, the strategy is to compare the change of astrometric data of celestial objects in a certain data collection interval. An ideal source for such data is \textit{Gaia} and \textit{Theia} measurements with observation time $T\sim 10$ yrs. In this scenario, the star term can be neglected. Starting from Eq. \ref{Eq:deltan} and defining $\delta\omega^i = \delta \dot{n}^i$, the equal-time two-point correlation of the proper motions due to a Gaussian isotropic stochastic background is given by \cite{book2011astrometric}
\begin{align}
    \Big\langle \delta \omega^i(\boldsymbol{n},t) & \delta \omega^j(\boldsymbol{n}',t) \Big\rangle = \frac{3}{8\pi} \Gamma_{ij,\bigoplus}(\Theta) \ \omega^2_{\rm rms}
\end{align}
where
\begin{align}
    \Gamma_{ij,\bigoplus}(\Theta)\equiv\sum_{A=\plus,\times}\Gamma^A_{ij,\bigoplus}(\Theta)
\end{align}
and
\begin{align}\label{omegarms}
    \omega^2_{\rm rms} \equiv H^2_0 \int^{f_{\text{high}}}_{f_{\text{low}}} d\ln f \ \Omega_{\rm mem}(f)
\end{align}
gives the rms of proper motions. Here, $\Omega_{\rm mem}(f) =\Omega_{\rm mem}(f,z_o=0)$ is the energy density per unit logarithmic interval in gravitational wave background due to the memory as of now, related to the strain power spectrum via Eq. \ref{Eq:strainomega} \cite{book2011astrometric}. Before going into the further details of how this effect can be analyzed in a given dataset, it is beneficial to make a comparison between non-memory and memory contributions to the SGWB, both of which contribute to the two-point correlation in the same way and with the same $\Gamma_{ij,\bigoplus}(\Theta)$. We note that for a SGWB sourced by SBBH mergers, 
\begin{align}\label{omegaintegral}
    &\Omega_{\text{non-mem}}(f) = A_0 (f/\text{yr}^{-1})^{2/3} \quad   f\leq f_{\rm high} \nonumber
    \\ &\Omega_{\text{mem}}(f) = \frac{2D(0)}{3H^2_0} f \qquad f_{\rm low}\leq f \leq f_{\rm high}
\end{align}
where $A_0 \simeq 10^{-9}$ (\cite{agazie2023nanograv}) and $D(0) = 10^{-38 \pm 2} \ \rm s^{-1}$ as shown in the $z_0 \rightarrow 0$ limit for $D(z_{0})$ for SBBH in Figure \ref{fig:D} (left). 

The low frequency cutoff is not relevant for this calculation since the integral of Eq. \ref{omegarms} is dominated by the high frequency cutoff. For the high frequency cut-off, we note that \textit{Gaia} outputs a single proper motion data averaged over $T$. Therefore, it acts as low-pass filter and motions above $2/T$ frequencies will be suppressed. Therefore, we take $f_{high}=2/T \sim 10^{-8}$ Hz same for non-memory and memory as well.

With these definitions, we proceed to evaluate Eq. \eqref{omegarms} using Eq. \eqref{omegaintegral} and obtain $\omega^2_{\rm rms} =  1.5 \times 10^{-7} (\mu\text{as/yr})^2$ for non-memory and $\omega^2_{\rm rms} = 2.8 \times 10^{-9\pm 2} (\mu\text{as/yr})^2$ for the memory. Again, we note that we consider only the SBBH model in this calculation, and that other GW contributions would further increase these estimates. 

We construct an estimator based on survey data, leveraging the above geometric correlation factor. Suppose we have $N_p$ number of celestial objects, $I=1,2,\dots N_p$, of which we have the data of proper motions: $\omega^{i}_{I}(t_p)$ with the corresponding error $\sigma_I(t_p)$, $p=1,2,\dots,N_t$. The data may be recorded as a function of time or could be averaged over an observation period $T$ to a single data point per object (which is the case for \textit{Gaia/Theia}). We then construct a likelihood function given by
\begin{align}\label{likelihood}
     \mathcal{L} \propto &\prod_{IJ,p} \exp\Big[-\big[\omega_{iI}(t_p) \omega_{jJ}(t_p)-\frac{3\widehat{\omega^2}}{8\pi^2} \Gamma_{ij,\bigoplus}(\theta_{IJ}) \big] \nonumber \\ & \times \big[\omega^{i}_I(t_p) \omega^j_{J}(t_p)-\frac{3\widehat{\omega^2}}{8\pi^2} \Gamma^{ij}_{\bigoplus}(\theta_{IJ})\big]/(2\sigma^2_{IJ,p})\Big]
\end{align}
Here, we clarify the notation. The Latin lower-case indices ($i,j$) are referring to the coordinate system on 2-sphere. The Latin upper-case indices ($I,J$) are referring to different celestial objects with specific directions $(\boldsymbol{n},\boldsymbol{n}')$ and $\cos{\Theta_{IJ}}=\boldsymbol{n} \cdot \boldsymbol{n}'$. Finally, $\widehat{\omega^2}$ is what we are trying to estimate (to be compared with Eq. \eqref{omegarms}) and $\sigma_{IJ,p}$ is denoting the error in two point correlation which we relate to $\omega^{i}_{I}(t_p)$ below.

To give an expression for $\sigma_{IJ,p}$ we start with
\begin{align}
    \sigma^2_{IJ,p} &= \Big\langle (\omega^{i}_{I}(t_p) \ \omega^{j}_{J} (t_p) \ \omega_{i,I}(t_p) \ \omega_{j,J}(t_p) \Big\rangle \nonumber \\&- \Big\langle \omega^{i}_{I}(t_p) \ \omega^{j}_{J}(t_p) \Big\rangle \Big\langle\omega_{i,I}(t_p) \  \omega_{j,J}(t_p)\Big\rangle
\end{align}
then utilizing Isserlis' theorem we write
\begin{align}
     \Big\langle(\omega^{i}_{I}(t_p) & \ \omega^{j}_{J}(t_q) \ \omega_{i,I}(t_p) \ \omega_{j,J}(t_q) \Big\rangle \nonumber \\ &= 2\Big\langle \omega^{i}_{I}(t_p) \ \omega^{j}_{J}(t_q) \Big\rangle \Big\langle \omega_{i,I}(t_p) \ \omega_{j,J}(t_q) \Big\rangle \nonumber \\&+ \Big\langle \omega^{i}_{I}(t_p)\ \omega_{i,I}(t_p) \Big\rangle \Big\langle \omega^{j}_{J}(t_q) \ \omega_{j,J}(t_q) \Big\rangle
\end{align}

We assume $\Big\langle \omega^{i}_{I}(t_p) \ \omega^{j}_{J}(t_q)\Big\rangle \ll \Big\langle \omega^{i}_{I}(t_p) \ \omega_{i,I}(t_p) \Big\rangle = \bar\sigma^2$, which holds when the observations are noise dominated and the noise is stationary and uncorrelated over different directions. Therefore, $\sigma^2_{IJ,p} \simeq \bar\sigma^4$. This relation allows one to relate estimate error of the Eq. \eqref{likelihood} in terms of the number of observed objects and $\bar\sigma$. Since the signal-to-noise ratio (SNR) is expected to scale as $\sqrt{N}/\sigma$ in general, we have
\begin{align}\label{sigma}
    \sigma_{\widehat{\omega^2}} \simeq \frac{\Bar{\sigma}^2}{\sqrt{N_t N_p(N_p-1)/2}}.
\end{align}

Table \ref{table:estimates} summarizes the predicted sensitivities for \textit{Gaia} and \textit{Theia} surveys. We assume for both surveys that the dataset consists of a single averaged proper motion measurement for each object over the observation time $T$, hence setting $N_t=1$. In particular, we estimate the sensitivities to $\widehat{\omega^2}$ at the level of $ 3.0 \times 10^{-2}$ \text{$\mu$as/yr} for \textit{Gaia} and  $ 3.0 \times 10^{-8}$ \text{$\mu$as/yr} for \textit{Theia}.

\renewcommand{\arraystretch}{1.5}
\begin{table}[h]
\centering
\caption{Proper motion uncertainties in \textit{Gaia/Theia} datasets for Quasi-Stellar Objects \protect\footnotemark. For \textit{Gaia} estimates, see Figure 6 in \cite{paine2018gaia} and for \textit{Theia} we use the estimates in \cite{jaraba2023stochastic}.}
\label{table:estimates}
\begin{tabular}{|l|l|l|}
\hline
 & \textit{\rm Gaia} & \textit{\rm Theia} \\ \hline
$N_p$  & $  10^6$  &  $ 10^8$ \\ \hline 
$\Bar{\sigma} \ (\text{$\mu$as/yr})$  & 200   & 2\\ \hline
$\sigma_{\widehat{\omega^2}} \  (\text{s}^{-2})$  & $ 6.6 \times 10^{-40} $  &  $ 6.6 \times 10^{-46}$ \\ \hline
$\sigma_{\widehat{\omega^2}} \ (\text{$\mu$as/yr})^2$  & $ 3.0 \times 10^{-2} $  &  $ 3.0 \times 10^{-8}$ \\ \hline
\end{tabular}
\end{table}

\footnotetext{The numbers $N_p$ here are smaller than the actual objects in these dataset. However, the full dataset has a larger $\Bar{\sigma}$ as well. Therefore, there is a optimum sub-class of the full dataset that would give the lowest smallest $\sigma_{\widehat{\omega^2}}$ which is what we are considering here.} 

These sensitivies are to be compared with the SBBH estimates of $1.5 \times 10^{-7} (\mu\text{as/yr})^2$ for non-memory and $\omega^2_{\rm rms} = 2.8 \times 10^{-9 \pm 2} (\mu\text{as/yr})^2$ for the memory. Therefore, the size of the effect we are considering here is potentially comparable to the non-memory part and can be potentially detected by \textit{Theia}.

Two comments are in order. First, the estimate above for \textit{Gaia} and \textit{Theia} assumes that all the sources have the same correlated root mean square motion. This is certainly not true for a realistic datasets. Second, the usual stochastic non-memory gravitational wave background contribution has the same angular correlation function as the memory contribution. Therefore, it is not possible to isolate the memory contribution just from the correlation analysis we did above. If the survey samples the peculiar motion many times $N_t\geq 1$ (instead of a single averaged measurement), it would be possible to analyze the two-point correlation in different frequency bins and hence distinguish the memory and non-memory contributions. In particular, at low frequencies the memory part starts dominating over the non-memory due to the different power indices in Eq. (\ref{omegaintegral}).

\subsection{Single snapshot analysis}
\label{sec:scenarioB}
At very low frequencies the star term ($\Omega^{\bigstar}_{\text{mem}} \sim 1/f$) becomes comparable to the Earth memory term ($\Omega_{\text{mem}} \sim f$) and cancels it out partially (Saturation Limit). This relation basically comes from the fact that star term is proportional to the anti-derivative of the Earth term which was given by $\Omega_{\text{mem}} = \frac{2D(0)f}{3H^2_0}$ and the fact that $\Omega(f) \propto h^2_c$. The second scenario outlined before consists of analyzing the very low frequency limit where the Earth and the star term partially cancel out and we are left with the 'Saturation Limit', $\delta n^{i}_{s}(\infty)$,  defined in Eq. \eqref{satdef}.

The single snapshot analysis consists of treating the observed angular distribution of the quasars in a CMB-like manner where the quasar density field is characterized by a mean, isotropic density plus a small anisotropic deviation from the mean. The statistics of the two point correlation of these fluctuations would depend on the galaxy clustering at the time of emission of light as well as propagation effects such as lensing and of course also on the memory accumulation we analyze in this paper. In lieu of a complete analysis which would take into account all possible sources of quasar clustering, we focus here on a simple estimation of the clustering due to memory and compare it to the expected shot noise due to the finite sample size (number of quasars) available. In this second scenario, since most astrometric deflections will be in the Saturation Limit, it is straightforward to predict the density pattern (as opposed to the deflection pattern since we only have access to a single snapshot). Using Eq. \eqref{satdef} we obtain
\begin{align}
    &\rho^{+} \equiv \boldsymbol{\triangledown} \cdot \delta \boldsymbol{n}^{+}_s(\infty) (\hat{z}) = -\frac{3}{2}A_0 \cos(2\phi) \sin^2(\theta)  \\\  &\rho^{\times} \equiv \boldsymbol{\triangledown} \cdot \delta \boldsymbol{n}^{\times}_s(\infty) (\hat{z})  = -\frac{3}{2} A_0 \sin(2\phi) \sin^2(\theta)
\end{align}
where the angles are standard spherical coordinates and we assumed a GW signal propagating in $\hat{z}$-direction. $A_0$ is the strength of a single GWM signal. In reality the over-density will be a sum of many such signals coming from different directions and the full effect will be of the order of the memory as estimated on right part of the Figure \ref{fig:D}, i.e. $A_0 \simeq h_c$.

\begin{figure}
\begin{minipage}{.5\textwidth}        \includegraphics[width=0.48\linewidth,angle=0]{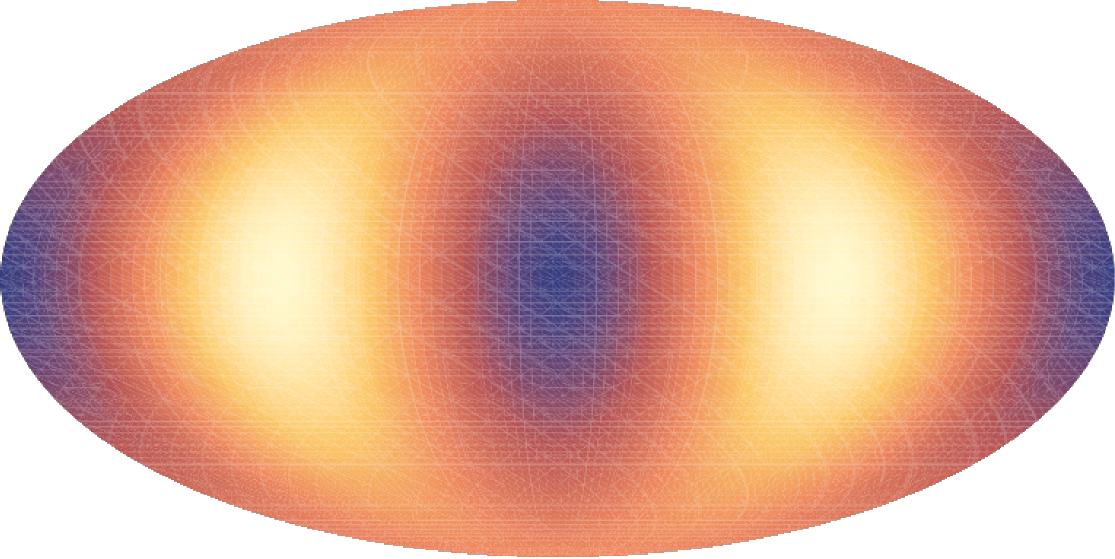}
\includegraphics[width=0.48\linewidth,angle=0]{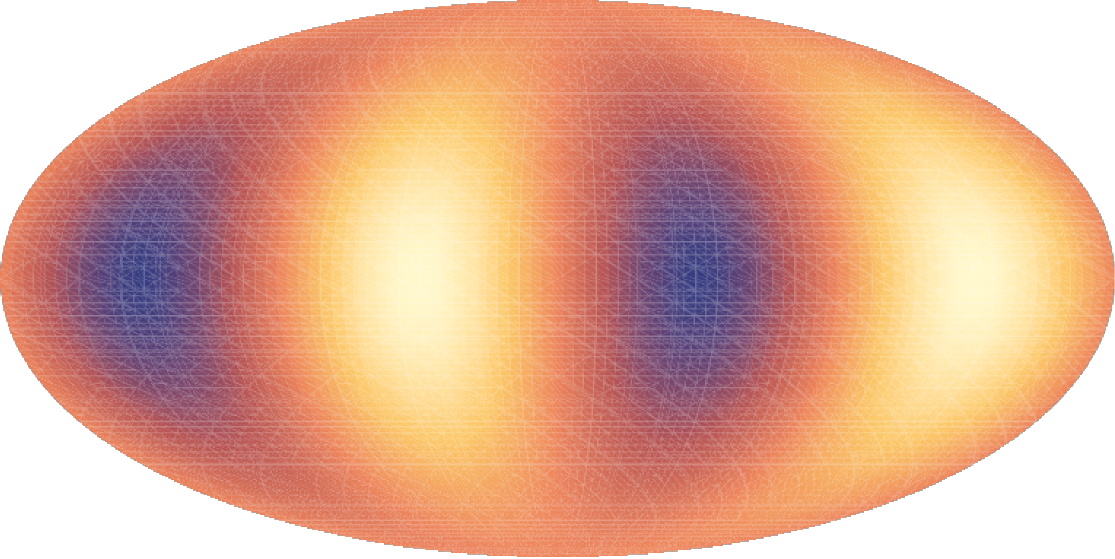}
\end{minipage}
\begin{minipage}{.5\textwidth}
\includegraphics[width=0.48\linewidth,angle=0]{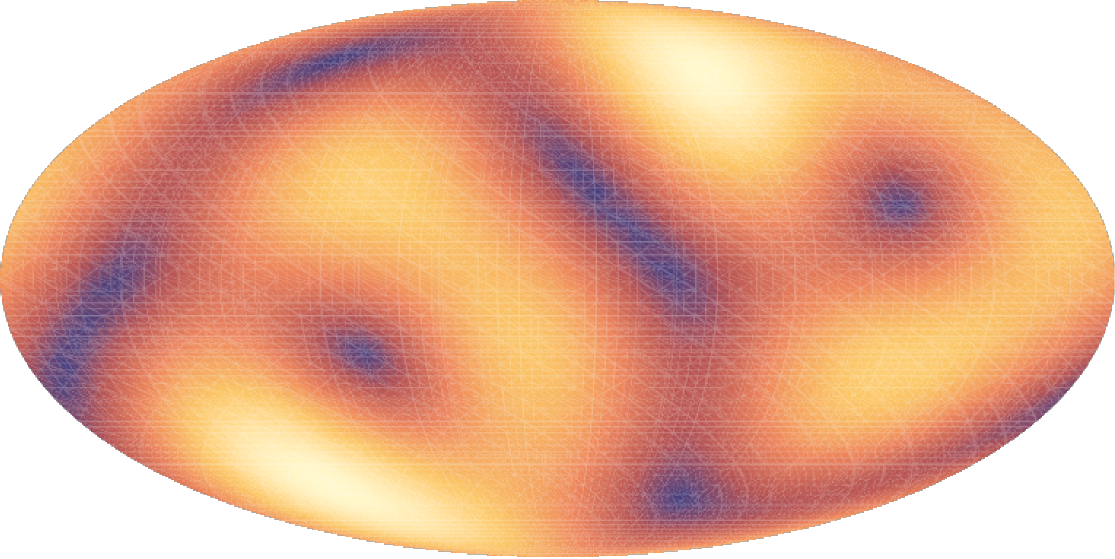}
\end{minipage}
    \caption{Top two figures depict density patterns of $+$ and $\times$ polarization of $\delta \boldsymbol{n}^{\plus,\times}_{s}$, respectively, for a GW propagating in the $\hat{z}$ direction. The bottom figure depicts a realization of the density pattern for many such GWs uniformly distributed in direction and polarization.}
    \label{fig:denssat}
\end{figure}

Assuming an isotropic initial distribution, we plot the correlation functions for this limit defined in Eq. \eqref{corrdef2} in the bottom row of Figure \ref{fig:corr}. We also plot a random realization of these density perturbations in the saturation limit in Figure \ref{fig:denssat}. The fact that there is a $ \rho \propto \sin^2(\theta)$ dependence on the right hand side implies that the astrometric deflections in the saturation limit are contributing only at the $l=2$ multipole level in the spherical harmonic expansion of the density fluctuation field. More specifically, expanding the smoothed quasar overdensity in spherical harmonics we get

\begin{eqnarray}
    \rho^A=a_{lm}^AY_{lm}(\theta,\phi)
\end{eqnarray}

where $\langle a_{lm}^A\rangle =0$ and $C_l=\langle |a_{l m}^+|^2+|a_{l m}^\times|^2\rangle$. The strain memory we estimated in the Section \ref{sec:results}, is around $10^{-(10-11)}$ for the dominant SBBH contribution. That implies a value for the $C_2$ coefficient of about $C_2\sim 10^{-(20-22)}$. The measurement of this coefficient is limited by the Poisson shot noise due to the limited sample size. For finite angular distributions the shot noise is $l$-independent and given by \cite{Campbell:2014mpa} $C_{l,{\rm shot}}=\frac{4\pi}{N}(1-\delta_{l0})$ which implies that one would need at least $10^{20}$ number of light sources to sufficiently suppress it. Since even the most ambitious astrometric observations such as \textit{Theia} have a projected sample size of about $10^9$ astronomical objects, we conclude that it is unlikely that such an anisotropy could be detected. Additionally, since the memory affects the low multipole moment, the measurement of $C_2$ is cosmic variance limited and hence, any other more significant effect that modifies the $C_2$ coefficient can not be subtracted from the theoretical expectation to a greater than ${\cal{O}} (1)$ precision. This makes the measurement of a subdominant effect, such as due to the memory, unlikely. For this reason our calculation can be understood as a proof of principle about the potential of the memory to affect the anisotropy of the universe, but realistically, an observable signature would require a much more powerful source of memory than the ones we study in the present work.

Before closing this section we would like to stress the limits of our calculation. Our implicit assumption of GW planar wavefronts implies that the GW sources are at a much greater redshift than the sources of light we consider. However, in the list of the presently observed quasars, most are at a distance of $z\sim {\cal O}(1)$ \cite{jaraba2023stochastic}. As a result, considering sources of GWs even further away would dramatically decrease the overall memory buildup. Instead, the better option would be to limit a potential analysis to the subgroup of quasars which are closer than $z\leq 1$ therefore allowing for a longer memory buildup while our approximation of planar wavefronts holds. This overall would imply a decrease of the total quasars available to $25\text{\%}$  which would decrease the result estimated above by an order of magnitude. Understanding the optimal trade-off between the number of quasars and duration of memory buildup or performing an improved calculation with spherical wavefronts as in \cite{book2011astrometric} is left for future work. 


\section{Discussion and Conclusions}
\label{sec:conclusions}
The memory effect is one of the most intriguing, as of now unconfirmed, predictions of general relativity. Unfortunately, direct detection efforts of the memory effect have been hampered so far due to the low frequency insensitivity of current ground based detectors. Alternative approaches that have been suggested for exploring the effects of memory focus on space-based detectors or astrometry. In either case, one is faced with the challenge of isolating the memory signal in face of a typically dominant primary GW signal. In this work, we explore whether one can leverage one of the fundamental differences between the primary and the memory signals to distinguish their phenomenology: namely, the primary signal is transient, whereas the memory signal is permanent. One would expect that over a long period of time, the small but permanent memory signals coming from different sources from different directions in the sky would add up in a random walk fashion and produce an effect that is ever increasing and potentially greater than even the present primary signals. We refer to this concept as `accumulating memory' and the purpose of this work was to explore the consequences of such an effect in the case of astrometry.

Even though in principle every possible GW source should contribute to the effect described above, we focused in this work on GWs coming from compact binary coalescences because these mergers produce some of the strongest GWs possible and also because their merger rates and mass distributions can be modeled even at high redshift, which is necessary for computing the GWM accumulation. Using this information, we provided the formalism for computing the accumulated GWM strain and derived the total GWM accumulation over the history of the universe from CBCs. 

Subsequently, we expanded upon the work of Madison \cite{Madison:2020xhh} on astrometric deflections from GWM by estimating the effect arising from many GW sources over the entire universe and separated the detection prospects into different scenarios: observations of the proper motions of distant objects, and a single snapshot observation of the angular distribution of very far away sources such as quasars. For the first case we found that the Earth term dominates and indicated that the angular correlation function of the proper motion of light sources should be the conventional one originally derived by Book \& Flanagan \cite{book2011astrometric}. Within this scenario we constructed memory SGWB sampled from realistic models of various binary populations and estimated the amplitude of apparent proper motions that would be induced by gravitational wave memory.
In the latter scenario we found that both the Earth and Star term (analogous to the Integrated Sachs-Wolfe effect for the CMB) contribute at the same level and the overall deflection would be dominated by GWs in the `Saturation Limit' which consists of the Earth and Star terms obtaining their maximum value. We derived the new angular correlation function for this limit and assessed the detection prospects with experiments such as \textit{Gaia} and \textit{Theia}. Our findings indicate that while it may be possible to detect the memory contribution to the proper motion of distant light sources, it is highly unlikely that the second scenario could be observationally confirmed.



%
\medskip\noindent\textit{Acknowledgments\,---\,}%
The authors would like to thank Colm Talbot for valuable conversations at an early part of this work. AP would like to thank Javier Carr\'on Duque for insightful comments. AP was supported by IBS under the project code, IBS-R018-D1. AP acknowledges support from the “Consolidación Investigadora” grant CNS2022-135590. AP’s work is partially supported by the Spanish Research Agency (Agencia Estatal de Investigación) through the Grant IFT Centro de Excelencia Severo Ochoa No CEX2020-001007-S, funded by MCIN/AEI/10.13039/501100011033. TB was in part supported by the McKnight Foundation. VM was in part supported by the NSF grant PHY-2110238. 

\newpage
\bibliographystyle{JHEP}
\bibliography{ref2}

\onecolumngrid

\end{document}